\documentstyle[12pt]{article}


\def\Lp{\displaystyle{\biggl(}}
\def\Rp{\displaystyle{\biggr)}}

\newcommand{\G}{\Gamma}
\newcommand{\D}{\Delta}

\renewcommand{\b}{\beta}
\renewcommand{\d}{\delta}
\newcommand{\e}{\varepsilon}

\renewcommand{\l}{\lambda} 
\newcommand{\m}{\mu}
\newcommand{\n}{\nu}

 \renewcommand{\O}{\Omega}

\newcommand{\s}{\sigma} \renewcommand{\S}{\Sigma}

\newcommand{\GG}{{\cal G}}

\newcommand{\SS}{{\cal S}}

\newcommand{\UU}{{\cal U}}

\newcommand{\complex}{{\kern .1em {\raise .47ex
\hbox {$\scriptscriptstyle |$}}
    \kern -.4em {\rm C}}}
\newcommand{\real}{{{\rm I} \kern -.19em {\rm R}}}
\newcommand{\rational}{{\kern .1em {\raise .47ex
\hbox{$\scripscriptstyle |$}}
    \kern -.35em {\rm Q}}}
\renewcommand{\natural}{{\vrule height 1.6ex width
.05em depth 0ex \kern -.35em {\rm N}}}
\newcommand{\dint}{\displaystyle{\int}}

\newcommand{\cb}{{\bar c}}

\newcommand{\pa}{\partial}

\newcommand{\dfrac}[2]{{\displaystyle{\frac{#1}{#2}}}}

\newcommand{\sla}{\raise.15ex\hbox{$/$}\kern -.57em}

\newcommand{\twiddle}{\lower.9ex\rlap{$\kern -.1em\scriptstyle\sim$}}

\newcommand{\equ}[1]{(\ref{#1})}

\newcommand{\eq}{\begin{equation}}
\newcommand{\eqn}[1]{\label{#1}\end{equation}}
\newcommand{\eea}{\end{eqnarray}}
\newcommand{\eqa}{\begin{eqnarray}}
\newcommand{\eqan}[1]{\label{#1}\end{eqnarray}}
\newcommand{\ba}{\begin{array}}
\newcommand{\ea}{\end{array}}
\newcommand{\eqac}{\begin{equation}\begin{array}{rcl}}
\newcommand{\eqacn}[1]{\end{array}\label{#1}\end{equation}}

\newcommand{\at}{{\~a}}

\newcommand{\ooo}{{\'o}}

\newcommand{\iii}{{\'\i}}

\begin{document}


{\ }

\vspace{3cm}
\centerline{\LARGE BRST Invariant mass term}\vspace{2mm}
\centerline{\LARGE for Chern-Simons}

\vspace{1cm}
\centerline{\bf {\large M.A.L.Capri, V.E.R.Lemes}}
\vspace{2mm}
\centerline{{\it UERJ - Universidade do Estado do Rio de Janeiro}}
\centerline{{\it Instituto de F{\iii}sica }}
\centerline{{\it Departamento de F{\iii}sica Te{\ooo}rica}}
\centerline{{\it Rua S{\at}o Francisco Xavier, 524}}
\centerline{{\it 20550-013 Maracan{\at}, Rio de Janeiro, Brazil}}
\vspace{4mm}


\centerline{{\normalsize {\bf REF. UERJ/DFT-03/99}} }

\vspace{4mm}
\vspace{10mm}

\centerline{\Large{\bf Abstract}}\vspace{2mm}
\noindent
A perturbative analysis of a massive Chern-Simons is presented.
The mass term is introduced in a BRST invariant way. With these method we prove
that the number of independent renormalisation of massive Chern-Simons
is equal to topologically massive Yang-Mills and different from pure Chern-Simons.
\newpage

\section{Introduction}
\noindent\hspace{0.75cm}In the last years, many efforts has been done in order
to extent the bosonization
mechanism to three dimensions \cite{w,3d,Nino,Gui}.
This efforts give rise to a large number of works that link planar effects that
appear in condensed matter physics
to vectorial actions. In particular the 2+1 dimensional Thirring model has been
bosonized into a free vectorial theory
in the leading order of the inverse mass expansion\cite{3d}, using the equivalence of the
abelian self-dual and the topologically massive model \cite{Mo}. This mechanism is extended to
prove that the computation of quantities that are mapped into a loop integral
are equivalent, to each oder by a field redefinition that is a power series, whose convergence depends on the relation of the energy and the mass parameter \cite{large,link,eg}.
Many works has been done recently  in order to extent this equivalence of
loop integrals into nonabelian three dimensional models \cite{Ruiz, Gia,bos,Ilha}.
One important point to emphasize here, is that all these results depends on the fact
that the number of independent renormalization parameters must be equal
in the two models, ie, the
bosonization process does not change the physical target space.

\noindent\hspace{0.75cm}This work has the purpose to prove that, from the point of view of the number of renormalization the three dimensional
Yang-Mills action is equivalent a massive Chern-Simons one. The method used to do this consists into introduce a insertion of mass dimension $2$ into the original Chern-Simons action as a BRST invariant term.
This is done with the use of a pair of fieds $\l, j$ that form a doublet including the mass as the expected value of the $j$ field in the BRST symmetry \cite{Ver,Dud,Sar}.

\noindent\hspace{0.75cm}The paper is organized as follows. In Sect.II we present the Chern-Simons action and introduce the mechanism that
turn possible to introduce mass into a invariant way. Sect.III is devoted to present all symmetries that are able to be extended
to quantum level and help us to fix the counterterm action. Sect.IV will be devoted to the study of the counterterms and the obtaining of the number of
independent renormalization constants presented into the model. Finally in the conclusions we discuss the results, comparing them with
the well know results obtained for Yang-Mills.

\section{The Chern-Simons action with mass term}
\noindent\hspace{0.75cm}In order to analyse the properties of the Chern-Simons action with mass term,
let us first begin by recalling the pure Chern-Simons action into the Landau gauge,
\eq
S = \dint d^{3}x \biggl[\dfrac{1}{4}\e^{\m\n\s} \biggl( A^{a}_{\s}F^{a}_{\m\n}
- \dfrac{1}{3}g f^{abc}A^{a}_{\m}A^{b}_{\n}A^{c}_{\s} \biggr) +
b^{a}\pa^{\m}A^{a}_{\m} + \cb^{a}\pa_{\m}(D^{\m}c)^{a} \biggr] .
\eqn{CS3}
\noindent\hspace{0.75cm}The expression \equ{CS3} is left invariant by the following nilpotent BRST transformations:
\eq\ba{ll}
 &s A^{a}_{\m} = -(D_{\m}c)^{a},\hspace{1.0cm} s c^{a}=\dfrac{1}{2}gf^{abc}c^{b}c^{c}\\
 &s \cb^{a} = b^{a},\hspace{2.35cm} s b^{a}=0.
\ea\eqn{brst}
\noindent\hspace{0.75cm}In order to study the renormalization properties of the Chern-Simons action plus a mass term.
Let us introduce these term into the action as a BRST variation of the type:
\eqa
s\Lp \dint d^{3}x\dfrac{1}{2}\l{A^{a}}_{\m}A^{a \m}\Rp = \dint d^{3}x \biggl[\dfrac{1}{2}(j+m)A^{a}_{\m}A^{a\m}+\l \pa_{\mu}c^{a}A^{a\mu}\biggr],
\eea
\noindent where $\l $ and $j$ form together a BRST doublet ($s\l = j + m$ and $s j=0$).
It is important to emphasize that these doublet is related to a product of gauge fields
and is equivalent to say that the spectrum of the gauge propagator is
changed introducing a mass pole.

\noindent\hspace{0.75cm}Now the complete fully quantized action with the invariant mass term and all Symmetries is given by:
\eq\ba{ll}
\S=\dint d^{3}x\biggl[\dfrac{1}{4}\e^{\m\n\s}\biggl( A^{a}_{\s}F^{a}_{\m\n}
- \dfrac{1}{3}g f^{abc}A^{a}_{\m}A^{b}_{\n}A^{c}_{\s}\biggr)
+ b^{a} \pa^{\m}A^{a}_{\m} + \cb^{a} \pa_{\m}(D^{\m}c)^{a}+\\\hspace{2.0cm}- \O^{a}_{\mu}(D^{\mu}c)^{a}
+\dfrac{1}{2}gf^{abc}L^{a}c^{b}c^{c}
+\dfrac{1}{2}(j+m)A^{a}_{\m}A^{a\m}+\l \pa_{\mu}c^{a}A^{a\mu}+\\
\hspace{2.0cm}+ \xi_{1}m^2j+\dfrac{\xi_{2}}{2} m j^{2}
+ \dfrac{\xi_{3}}{6}  j^{3}\biggr]
\ea\eqn{sigma}
\noindent The last three terms that appear in the action are only to complete the most general power-counting invariant case.

\section{Ward identities}

\noindent\hspace{0.75cm}Now we must collect all symmetries that could be writen into a form of a ward identity.
These Symmetries that are compatible with the quantum action principle \cite{livro}give rise
to a complete set of equations that defines in a unique way all the
possible counterterms that are permited for these action. These ward identities are given by:

\noindent -The equation that defines the insertion,
\eq
\UU(\S ) = \int d^{3}x \Lp \dfrac{\d \S}{\d \l} + c^{a}\dfrac{\d \S}{\d b^{a}} \Rp =0;
\eqn{insert}
\noindent -The gauge fixing and antighost equation,
\eq
\dfrac{\d \S}{\d b^{a}} = \pa^{\mu}A^{a}_{\m},\qquad\dfrac{\d \S}{\d \cb^{a}} + \pa^{\m}\dfrac{\d \S}{\d \O^{ a \, \m}}=0;
\eqn{gf-aght}
\noindent -The ghost equation,
\eq\ba{ll}
\hspace{5cm}\GG^{a} \S = - \D^{a},\nonumber \\  \nonumber \\
\GG^{a} = \dint d^{3}x \Lp \dfrac{\d\enskip }{\d c^{a}} + gf^{abc} \cb^{b}\dfrac{\d\enskip }{\d b^{c}}  \Rp\qquad
\D^{a} = \dint d^{3}x[gf^{abc}(A^{b}_{\m}\O^{c \, \m} + L^{b}c^{c})];
\ea\eqn{def-ghost}
\noindent -The SL(2R)
\eq
D(\S ) = \int d^{3}x \Lp c^{a}\dfrac{\d \S}{\d \cb^{a}} + \dfrac{\d \S}{\d L^{a}}\dfrac{\d \S}{\d b^{a}} \Rp =0;
\eqn{sl2r}
\noindent -The Slavnov Taylor equation,
\eq
\SS (\S ) = \int d^{3}x \Lp \dfrac{\d \S}{\d A^{a}_{\m}}\dfrac{\d \S}{\d \O^{a \, \m}}
+ \dfrac{\d \S}{\d c^{a}}\dfrac{\d \S}{\d L^{a}} + b^{a}\dfrac{\d \S}{\d \cb^{a}}
+ (j + m)\dfrac{\d \S}{\d \l}  \Rp = 0.
\eqn{slavnov}
\noindent\hspace{0.75cm}In this point is important to emphasize that the lack of the usual vectorial symmetry,
presented in pure Chern-Simons\cite{livro}, that generates the algebra of supersymmetry is due to
the insertion term in the action. This insertion change the behaviour of the gauge propagator introducing one scale in the original Chern-Simons action.
The mass parameter introduced by the doublet $\l, j$ in the action change the cohomology
of the BRST operator \cite{livro} acting on the space of all fields
and sources. Into more simple words, in spite of coming from a BRST doublet, the mass
term change the behavior of the gauge propagator from topological to a metrical one.

\noindent\hspace{0.75cm}Now let us display for further use, the quantum numbers of all fields and sources entering in the action.
\begin{table}[htbp]
\begin{center}
\begin{tabular}{|l|l|l|l|l|l|l|l|l|}
\hline
Fields and souces & $A_\mu ^a$ &  $c^a$ & $\,\bar c^a$  & $b^a$ & $\,\,\lambda$ & $j$ & $\Omega^a_\mu$ & $L^a$ \\
\hline
Gh. number & \thinspace 0 & \thinspace 1 & $-1$ & 0 & $-1$
& 0 & $-1$ & $-2$ \\
\hline
Dimension & \thinspace 1 & $\,\,$0 & $\,$1 & $\,$1 & $\,\,$1 & 1 & $\,$2 &
$\,3$ \\
\hline
\end{tabular}\\
\caption{Ghost number and canonical dimension of the fields and souces.}
\label{tab.1}
\end{center}
\end{table}
\vspace{-1cm}
\section{Characterization of the general local invariant counterterm}

\noindent\hspace{0.75cm}In order to characterize the general invariant counterterm that is freely adeed to all orders of perturbation theory,
we perturb the classical action $\S $ by adding an arbitrary integrated local polinomial
in the fields and external sources $\G_{c}$ that has dimension limited to 3 and ghost number zero.
The requirement that the perturbed action obeys the same set of equation obtained for $\S $ give rise to the
full set of constraints under the quantum action $\G $:
\eqa
&\SS (\G) = 0,\qquad \UU(\G ) =0,\qquad \GG^{a}(\G) = -\D^{a}\cdot\G,\nonumber\\
&D(\G)=0,\qquad\dfrac{\d \G}{\d b^{a}} = \pa^{\mu}{A_{\mu}}^{a},\qquad\dfrac{\d \G}{\d \cb^{a}} + \pa^{\m}\dfrac{\d \G}{\d \O^{ a \, \m}} = 0.\label{esta1}
\eea
\noindent\hspace{0.75cm}The calculation of the Counterterm is obtined using
\eq
\G = \S + \epsilon \G_{c}
\eqn{def-contra}
\eqa
&\b_\S(\G_c)=0,\qquad \UU(\G_c ) =0,\qquad \GG^{a}(\G_c) =0,\nonumber\\
&D(\G_c)=0,\qquad\dfrac{\d \G_c}{\d b^{a}} =0,\qquad\dfrac{\d \G_c}{\d \cb^{a}} + \pa^{\m}\dfrac{\d \G_c}{\d \O^{ a \, \m}} = 0,\label{esta2}
\eea
\noindent where $\b_{\S}$ is the linearized nilpotent operator that extends the BRST symmetrie to all fields and sources and it is given by:
\eq
\b_{\S} = \dint d^{3}x \Lp \dfrac{\d\S}{\d A^{a}_{\m}}\dfrac{\d\quad }{\d \O^{a\m}}
+ \dfrac{\d \S}{\d \O^{a \, \m}}\dfrac{\d\quad }{\d A^{a}_{\m}}
+ \dfrac{\d \S}{\d c^{a}}\dfrac{\d\quad }{\d L^{a}} + \dfrac{\d \S}{\d L^{a}}\dfrac{\d\enskip }{\d c^{a}} + b^{a}\dfrac{\d\enskip }{\d \cb^{a}}
+ (j + m)\dfrac{\d \enskip}{\d \l}\Rp.
\eqn{linearizado}
\noindent\hspace{0.75cm}The Counterterm is obtined by the most general functional of dimension 3 and ghost number zero,
that obeys the equation $\b_{\S}(\G_{c})=0$, and it is given by:
\eq
\G_{c} = \dint d^{3}x\biggl[\dfrac{y}{4}\e^{\m\n\s}\biggl( A^{a}_{\s}F^{a}_{\m\n}
- \dfrac{1}{3} gf^{abc}A^{a}_{\m}A^{b}_{\n}A^{c}_{\s} \biggr)\biggr] + \b_{\S} \D^{(-1)}.
\eqn{forma-count}
\noindent According to the Table \ref{tab.1},
the most general form of $\D^{(-1)} $, that is a field polinomial of dimension $3$ and ghost number $-1$, is given by:
\eq\ba{ll}
\D^{(-1)} = \dint d^{3}x\biggl[a_{1} ({\O^{a}}_{\m} + \pa_{\m}\bar{c}^{a})A^{a\m}+a_{2}L^{a}c^{a}+ a_{3} \dfrac{1}{2}m gf^{abc}\bar{c}^{a}\bar{c}^{b}c^{c}+\l\biggl(a_{4}\dfrac{1}{2}A^{a\m}A^{a}_{\m}+\\
\hspace{2.7cm}+a_{5}\dfrac{1}{2}mj + a_{6}\dfrac{1}{6}j^{2}+a_{7}m\bar{c}^{a}c^a + a_{8}j\bar{c}^{a}c^{a}+a_9\bar c^ab^a+a_{10}b^ab^a\biggr)\biggr].
\ea\eqn{delta-1}
\noindent Using the full set of constraints given by \equ{esta2}, we have for the functional $\D^{(-1)}$:
\eq
\D^{(-1)} = \dint d^{3}x \biggl[ a_{1} ({\O^{a}}_{\m} + \pa_{\m}\bar{c}^{a})A^{a\m} - a_{1}\dfrac{1}{2}\l A^{a \, \m}A^{a}_{\m}
+ a_{5}\dfrac{1}{2}m\l j + a_{6}\dfrac{\l}{6}j^{2} \biggr].
\eqn{delta-f}
\noindent Then the term $\b_\S\D^{(-1)}$ in \equ{forma-count} becomes
\eq\ba{ll}
\b_\S\D^{(-1)}=\dint d^3x\biggl[a_1\biggl(\dfrac{1}{2}\e^{\m\n\s}A^a_\s F^a_{\m\n}+\dfrac{1}{2}(j+m)A^{a\m}A^a_\m+2\l\pa_\m c^aA^{a\m}+\\
\hspace{3.2cm}+(\O^a_\m+\pa_\m\bar c^a)\pa^\m c^a\biggr)+\dfrac{a_5}{2}(j+m)mj+\dfrac{a_6}{6}(j+m)j^2\biggr],
\ea\eqn{nao-sei}
\noindent and the counterterm action becomes:
\eq\ba{ll}
\G_c=\dint d^3x\biggl[\dfrac{y+2a_1}{4}\e^{\m\n\s}A^a_\s F^a_{\m\n}-\dfrac{y}{12}g\e^{\m\n\s}f^{abc}A^a_\m A^b_\n A^c_\s+a_1(\O^a_\m+\pa_\m\bar c^a)\pa^\m c^a+\\
\hspace{1cm}+\dfrac{a_1}{2}(j+m)A^{a\m}A^a_\m+2a_1\l\pa_\m c^aA^{a\m}+\dfrac{a_5}{2}m^2j +\Bigl(\dfrac{a_5}{2}+\dfrac{a_6}{6}\Bigr)mj^2+\dfrac{a_6}{6}j^3\biggr].
\ea\eqn{count-act}
\noindent\hspace{0.75cm}The 4 free parameters $(y, a_{1}, a_{5}, a_{6})$ present in the counterterm action are responsible
for the renormalization of all fields, sources the coupling constant $g$ and parameters $\xi_{i}$. Indeed is immediate to observe that
 \eq
\S + \epsilon\G_{c} = \S(A_{0},c_{0},\bar{c}_{0},b_0,\l_{0},j_{0},\O_{0},L_{0},\xi_{0},m_{0}, g_{0}) + O(\e^{2}),
\eqn{def-all}
\noindent by redefining the couplings, sources and the field amplitudes according to
\eq\ba{ll}
\varphi_0=Z_\varphi\varphi,\\
Z_\varphi=1+\epsilon z_\varphi
\ea\eqn{Z}
\noindent where $\varphi\equiv A^a_\m,c^a,\bar c^a,b^a,\l,j,\O^a_\m,L^a,m,g,\xi_1,\xi_2,\xi_3$. Using \equ{Z} in \equ{def-all} we have,
\eq\ba{ll}
Z_g =1+\epsilon z_g=1-\epsilon \dfrac{y}{2},\vspace{0.05cm}\\
Z_A =1+\epsilon z_A=1+\epsilon \biggl(\dfrac{y}{2} + a_{1}\biggr),\vspace{0.05cm}\\
Z_{m} =1+\epsilon z_{m},\vspace{0.05cm}\\
Z_c =1+\epsilon z_c=1-\epsilon \dfrac{a_1}{2},\vspace{0.15cm}\\
Z_{\overline{c}} =1+\epsilon z_{\overline{c}}=1-\epsilon\dfrac{a_1}{2},\vspace{0.05cm}\\
Z_{L} =1+\epsilon z_{L} =1+\epsilon \biggl(\dfrac{y}{2} + a_{1}\biggr),\vspace{0.05cm}\\
Z_{\l} =1+\epsilon z_{\l} =1-\epsilon \biggl(\dfrac{y}{2} - \dfrac{3a_{1}}{2}\biggr),\vspace{0.05cm}\\
Z_{b} =1+\epsilon z_{b} = 1-\epsilon\biggl(\dfrac{y}{2} + a_{1}\biggr),\vspace{0.05cm}\\
Z_{j} =1+\epsilon z_{j} =1+\epsilon z_{m},\vspace{0.05cm}\\
Z_{\xi_{1}}  =1+\epsilon z_{\xi_{1}} =1+\epsilon\biggl(\dfrac{a_{5}}{2} - 3 z_{m}\biggr),\vspace{0.05cm}\\
Z_{\xi_{2}}  =1+\epsilon z_{\xi_{2}} =1+\epsilon \biggl(\dfrac{a_{5}}{2} -\dfrac{a_{6}}{3} - 3 z_{m}\biggr),\vspace{0.05cm}\\
Z_{\xi_{3}}  =1+\epsilon z_{\xi_{3}} =1+\epsilon (a_{6} - 3 z_{m}).
\ea\eqn{Zs2}
In terms of the renormalization constants $Z_\varphi$. We have the following relations:
\eq\ba{ll}
Z_{\overline{c}} = Z_{c} = Z_{\O}={(Z_g)}^{-1/2}{(Z_A)}^{-1/2},   \\
Z_{L}= Z_{A},  \\
Z_{b} = {(Z_A)}^{-1},  \\
Z_{\l} = {(Z_g)}^{5/2}{(Z_A)}^{3/2}, \\
Z_{j} = Z_{m}.  \\
\ea\eqn{Zs3}
\noindent\hspace{0.75cm} We note from \equ{Zs2} that there are 3 physical independent
renormalizations $Z_g,Z_A,Z_m$, these renormalizations corresponds to the beta function, the anomalous dimension and renormalization of the mass pole. Is important to emphasize here that the mass introduced, in spite of coming from a BRST invariant term, change the propagator introducing one scale in the action. This scale is responsible for the breaking the supersymmetric structure presented in pure Chern-Simons \cite{livro} saying to us that the action now is not topological any more and has metric observables.   The parametrs $\xi_{1},\xi_{2},\xi_{3}$ are introduced only by consistency with the power counting and do not correspond to real physical parameters, all of the $ \xi_{i}$ can be set to zero at the end of the calculations. We see therefore that the number of independent physical renormalizations that appear in
massive Chern-Simons is equal to the number obtained in the case of  topologically massive Yang-Mills \cite{livro, sasaki, Chen}.
In these sense the duality mechanisms between these two actions could be implented without changing the number of  renormalization constants presented in topologically massive Yang-Mills or in massive Chern-Simons.
\section{Conclusion}
\noindent\hspace{0.75cm}We have presented a BRST analysis of Chern-Simons with a mass term.
This mass is introduced into a BRST invariant way, with the help of a BRST
doublet $\l $ and $j$.
These way of introduce mass, change the number and the form of the equations compatible with the quantum
action principle. The change in the pole of the propagator, from a massless to a massive one and the lack of the vectorial symmetry, change the behavior of the theory from topological to a metrical one. The two
new ward identities, the defintion of the mass insertion \equ{insert} and the SL(2R) ones\equ{sl2r}, plays a crucial role into the proof of the stability of the theory under radiative corrections and which turns out to have
important consequences for fixing the number of independent renormalizations and consequently
the metric behavior of the observables.

\noindent\hspace{0.75cm}Let us conclude by saying that these method of include insertions that change the propagator of a given action, making use of the insertions like $A^{2}$, could be very usefull
to study the duality equivalences under three dimensional  gauge actions, the relations under three dimensional actions by a duality mechanism using this insertion method, will be presented in future works.

\vspace{.5cm}
\noindent
{\large\bf{Acknowledgements}}
\newline
The Coordena{\c{c}}{\~{a}}o de Aperfei{\c{c}}oamento de Pessoal de N\'{\i}vel
Superior (CAPES) and the SR2-UERJ are acknowledged for the financial support.

\end{document}